\documentclass[floats,floatfix,amssymb,prl,twocolumn,superscriptaddress,nofootinbib]{revtex4-1}
		
\usepackage{amssymb,amsmath,verbatim,mathtools,needspace,enumitem,etoolbox,graphicx,physics,microtype,afterpage,bm}
\usepackage[dvipsnames, usenames]{xcolor}
\definecolor{linkcolor}{rgb}{0.0,0.3,0.5}
\usepackage[unicode, colorlinks=true, linkcolor=linkcolor, citecolor=linkcolor, filecolor=linkcolor,urlcolor=linkcolor, pdfusetitle]{hyperref}
\usepackage[all]{hypcap}
\usepackage[T1]{fontenc}
\usepackage[utf8]{inputenc}
\usepackage{tabularx}
\usepackage{float}
\usepackage{lmodern}
\allowdisplaybreaks
\usepackage{tikz}
\usepackage{color}
\usepackage{framed}
\usepackage{hyperref}
\hypersetup{colorlinks, citecolor=bluscuro, linkcolor=black, urlcolor=bluscuro}
\definecolor{rossos}{cmyk}{0,1,1,0.55}
\definecolor{bluscuro}{rgb}{0.15, 0.2, .85}
\definecolor{bluchiaro}{cmyk}{1,.3,0.,0.1}
\definecolor{ForestGreen}{rgb}{0.13, 0.55, 0.13}

\newcommand{\be}{\begin{equation}}
\newcommand{\ee}{\end{equation}}
\newcommand{\bea}{\begin{equation}\begin{aligned}} 
\newcommand{\eea}{\end{aligned}\end{equation}}
\renewcommand{\d}{{\rm d}}

\newcommand{\lp}{\left (}
\newcommand{\rp}{\right )}

\def\lsim{\mathrel{\rlap{\lower4pt\hbox{\hskip0.5pt$\sim$}}
    \raise1pt\hbox{$<$}}}         
\def\gsim{\mathrel{\rlap{\lower4pt\hbox{\hskip0.5pt$\sim$}}
    \raise1pt\hbox{$>$}}}         

\def\d{{\rm d}}

\def\SGWB{\text{\tiny SGWB}}
\def\SMBH{\text{\tiny SMBH}}
\def\HPBH{\text{\tiny  HPBH}}
\def\PBH{\text{\tiny PBH}}
\def\rad{\text{\tiny rad}}
\def\eq{\text{\tiny eq}}
\def\DM{\text{\tiny DM}}
\def\cl{\text{\tiny cl}}

\definecolor{myforestgreen}{rgb}{0.13, 0.55, 0.13}

\begin{document}

\title{Heavy primordial black holes from strongly clustered light black holes}

\author{Valerio De Luca}
\email{vdeluca@sas.upenn.edu}
\affiliation{Center for Particle Cosmology, Department of Physics and Astronomy,
University of Pennsylvania 209 S. 33rd St., Philadelphia, PA 19104, USA}

\author{Gabriele Franciolini}
 \email{gabriele.franciolini@uniroma1.it}
 \affiliation{Dipartimento di Fisica, Sapienza Università 
 	di Roma, Piazzale Aldo Moro 5, 00185, Roma, Italy}
 \affiliation{INFN, Sezione di Roma, Piazzale Aldo Moro 2, 00185, Roma, Italy}

\author{Antonio Riotto}
\email{antonio.riotto@unige.ch}
\affiliation{D\'epartement de Physique Th\'eorique and Gravitational Wave Science Center (GWSC), Universit\'e de Gen\`eve, CH-1211 Geneva, Switzerland CH-1211 Geneva, Switzerland}

\begin{abstract}
\noindent
We show that heavy primordial black holes may originate from much lighter ones if the latter are strongly clustered at the time of their formation. 
While this population is subject to the usual constraints from late-time universe observations, its relation to the initial conditions is different from the standard scenario and provides a new mechanism to generate massive primordial black holes even in the absence of efficient accretion,  opening new scenarios, e.g. for the generation of supermassive black holes.
\end{abstract}

\maketitle

\noindent{{\bf{\it Introduction.}}}
Multiple detections of gravitational waves (GWs) 
coming from black hole binary mergers~\cite{LIGOScientific:2016aoc, LIGOScientific:2018mvr,LIGOScientific:2020ibl, LIGOScientific:2021djp} have revived the interest in the 
physics of Primordial Black Holes (PBHs)~\cite{Sasaki:2018dmp, Carr:2020gox, Green:2020jor,Franciolini:2021nvv}. Indeed, some of the LIGO/Virgo/KAGRA data may be of primordial origin \cite{Bird:2016dcv,Sasaki:2016jop,Clesse:2016vqa,Ali-Haimoud:2017rtz,Hutsi:2020sol, DeLuca:2021wjr,Franciolini:2021tla,Franciolini:2022tfm} 
and future GW experiments will shed light on the possible existence of PBHs \cite{Chen:2019irf,Pujolas:2021yaw,DeLuca:2021hde,Bavera:2021wmw,Franciolini:2022htd,Cole:2022ucw}.

PBHs in the early universe are commonly born in the radiation-dominated phase (see Ref.~\cite{Green:2020jor} for a review on the various formation mechanisms). Given our ignorance of the production mechanism giving rise to PBHs, if any, we do not know if they are born randomly distributed or with a strong correlation among them.
In the standard scenario where PBHs are generated by the collapse of large overdensities created during inflation on small scales~\cite{Sasaki:2018dmp}, PBHs are Poisson distributed in space~\cite{Desjacques:2018wuu,Ali-Haimoud:2018dau,Ballesteros:2018swv,MoradinezhadDizgah:2019wjf}. However, large initial correlations are conceivable and not ruled out, unless the PBH abundance in the universe (normalised to the dark matter) $f_\PBH$ is larger than $\mathcal{O}(0.1)$ in the stellar mass range~\cite{DeLuca:2022uvz}.

In this Letter we propose a new mechanism to generate heavy PBHs from light ones making use of the large initial PBH clustering, a mechanism that we dub ``clusteringenesis''. The idea is quite simple: if PBHs are born close to each other, the strong gravitational interactions among them may result in the collapse of this clump into a more massive PBH, even in the radiation dominated phase of the early universe. This mechanism provides a novel way to increase the mass of light PBHs in the primordial epochs, which is alternative to the more standard process relying on baryonic mass accretion, which is efficient only for PBHs with mass larger than $\mathcal{O}(10) M_\odot$~\cite{DeLuca:2020qqa} at lower redshifts.

In the following we will describe the basics of this idea and discuss some of its possible implications.

\vskip 0.5cm
\noindent
\noindent{{\bf{\it Collapse of large overdensities  in the  radiation-dominated era.}}} 
Independently from the formation mechanism and being discrete objects, the most generic initial two-point correlator for the PBH density contrast $\delta_\PBH = \delta \rho_\PBH/\rho_\PBH$ acquires the form~\cite{Desjacques:2018wuu}
\be
\langle \delta_\PBH(\vec{r})\delta_\PBH(0)\rangle=\frac{1}{\overline{n}_\PBH}\delta_\text{\tiny D}(r)+\xi_\PBH(r),
\ee
in terms of their distance $r$, where 
\be
\overline{n}_\PBH \simeq 30 f_\PBH \lp \frac{M_\PBH}{M_\odot} \rp^{-1} {\rm kpc}^{-3}
\ee
is the average PBH number density per comoving volume for a monochromatic PBH population with mass $M_\PBH$.

We suppose that, at the time of formation, such two-point correlator is  dominated by the reduced correlation function $\xi_\PBH(r)$ up to some  comoving clustering scale $r_\cl$, while on  larger scales   the Poisson shot noise,  arising from the discrete nature of PBHs, dominates. For simplicity and for the sake of the argument,  we  assume an approximately constant in space and large reduced two-point correlation function up to $r_\cl$,
\be
    \xi_\PBH(r)\simeq
\begin{cases}
    \xi_0 \gg 1 \qquad {\rm for} \qquad r\lsim r_\cl,  \\
    0 \qquad \qquad \ \, {\rm otherwise}.
\end{cases}
\ee
We will be agnostic in the following regarding the origin of such correlations. Let us just point out that  large clustering appears if the local properties of the PBH overdensity field are space dependent. This effect might either come from an actual field different from the overdensity field, or from a long wavelength modulation of the overdensity field itself, resulting from a self-coupling of long and short scales as happens, e.g., in local models of non-Gaussianity~\cite{Atal:2020igj}. Alternatively, large correlation may arise if PBHs are formed by bubble collisions in first-order phase transitions~\cite{Khlopov:1998nm,Jung:2021mku,DeLuca:2021mlh} or for PBHs generated thanks to long-range scalar forces~\cite{Flores:2020drq}.
These models can be used to construct explicit realisations of the scenario proposed in this work.
Once rescaled to the total dark matter density $\delta \rho_\PBH/\rho_\DM \equiv \Phi \simeq \xi_0 f_\PBH$, the corresponding density contrast is suppressed by the PBH abundance, which we will assume to be tiny in the following in order to avoid bounds coming from CMB anisotropies on isocurvature perturbations~\cite{Planck:2018vyg, Young:2015kda, DeLuca:2021hcf}. 

 \begin{figure*}[t!]
	\centering
	\includegraphics[width=0.99\textwidth]{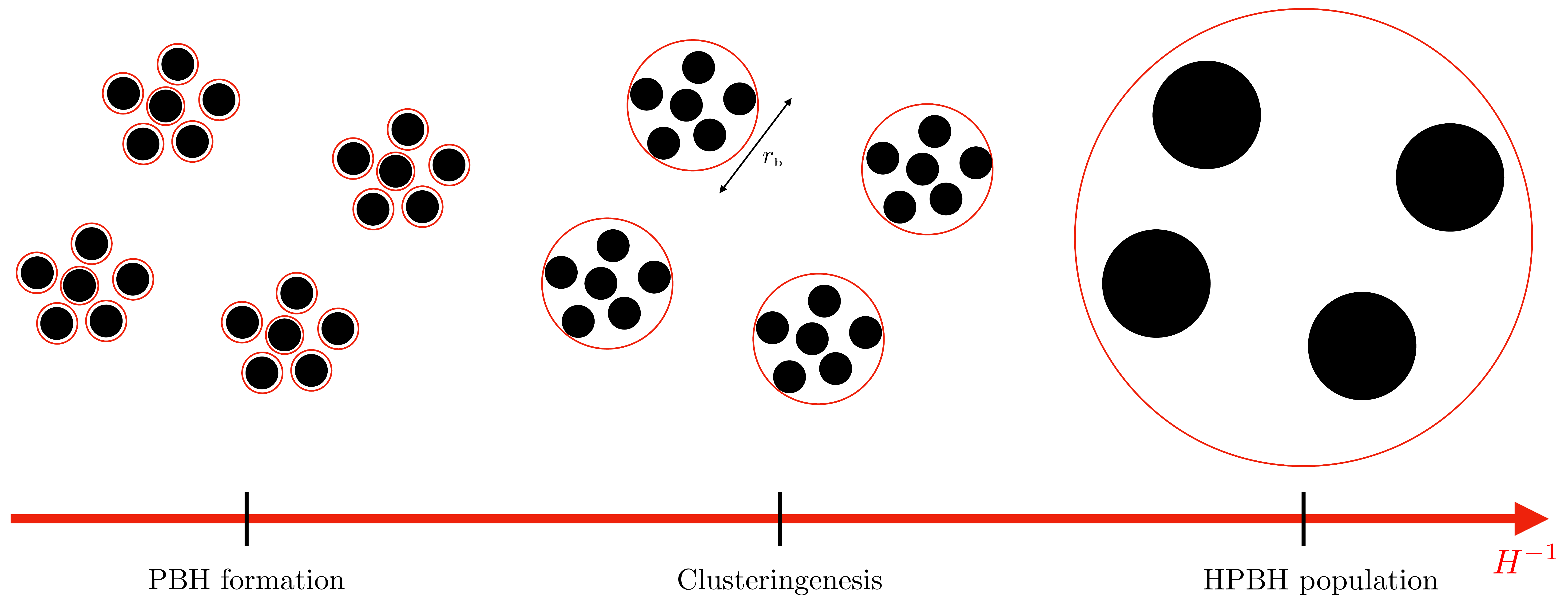}
	\caption{Pictorial representation of the clusteringenesis mechanism. The red line denotes the growing cosmological horizon $H^{-1}$.}
	\label{picture}
\end{figure*} 

The key  point is the subsequent evolution of such  large density PBH clumps during the radiation phase.  The common lore is that they do not grow till the  matter-radiation equality epoch is reached, because of the counteraction of the radiation pressure~\cite{Meszaros:1974tb}. However, this is true only at the linear level. 
In the full non-linear theory, the self gravity of these large non-linear fluctuations may become important before the equality time, and consequently give rise to their collapse and production of 
very dense clusters, after they decouple from the general expansion and virialize~\cite{Kolb:1994fi}.

Adopting the spherical collapse model, one can write down the evolution equation for the  parameter $R$, describing the deviation of the motion of each collapsing shell from the uniform Hubble flow of the background Friedmann universe~\cite{Kolb:1994fi}
\be
x (1+x) \frac{\d^2 R}{\d x^2} + \lp 1+\frac{3}{2}x\rp \frac{\d R}{\d x} + \frac{1}{2}\lp \frac{1+\Phi}{R^2}-R\rp = 0,
\ee
as a function of the rescaled scale factor $x = a/a_{\eq}$, in terms of the 
one at the epoch of matter-radiation equality $a_{\eq}$.
This equation assumes that the statistical distribution of the initial overdensities (or clumps) is determined by the correlation function $\xi_\PBH$; see the appendices of Ref.~\cite{DeLuca:2022uvz} for details on the relation between the profile of initial clumps and the correlation function.
In particular, at scales smaller than the clump separation, 
the density profile of the clusters is directly related to the correlation function and its evolution depends on its amplitude $\xi_0$~\cite{1999coph.book.....P}.

An analytic approximation for the solution can be obtained with a power expansion in the rescaled scale factor $x$, giving \cite{Kolb:1994fi}
\be
R \simeq 1 - \frac{\Phi x}{2} - \frac{\Phi^2 x^2}{8} + \mathcal{O} (x^3),
\ee
from which  one can show that the decoupling occurs approximately at $a_{\cl} = a_{\eq}/\Phi$. The corresponding timescale is approximately given by the free-fall time~\cite{1987gady.book.....B} 
\be
\label{freefall}
\tau_{\cl}= \sqrt{\frac{3\pi}{32 G \rho_\cl}} \simeq 1.2 \cdot 10^4 \Phi^{-2} \lp \frac{C}{200} \rp^{-1/2} {\rm yr},
\ee
expressed in terms of the average density of such clusters after relaxation $\rho_{\cl} = C \rho_{\rad} (a_{\cl}) = C \rho_{\eq} \Phi^4$~\cite{Kolb:1993zz,Kolb:1994fi}, where $C = {\cal{O}}(1 \divisionsymbol 10^2)$ is a constant that describes the overdensity amplitude. Their mass and physical radius are given by~\cite{DeLuca:2022uvz}
\begin{align}
 M_\cl  & \simeq 1.3 \cdot 10^2 \Phi \lp \frac{r_\text{\tiny cl}}{\rm kpc}\rp^3 M_\odot,\nonumber \\
   r_\text{\tiny b} &\simeq 4 \cdot 10^{-5} 
    \Phi^{-1}
    \lp \frac{C}{200} \rp^{-1/3}
    \lp \frac{r_\text{\tiny cl}}{\rm kpc} \rp {\rm kpc}.
\end{align}
Notice that the mass of the cluster $M_\cl$ does not depend on the individual PBH mass $M_\PBH$ because of two competing effects that cancel out. 
Indeed, for a given correlation function, 
a fixed PBH abundance may either result into heavier PBHs
with a smaller number density, or into lighter and more abundant ones.

\vskip 0.5cm
\noindent
\noindent{{\bf{\it Formation of heavy PBHs by initial clustering.}}} 
The key point of this Letter is that the  PBH clusters may collapse into  PBHs of mass $\approx M_\text{\tiny cl}$ if the final halo is more compact than a BH, giving rise to a population of heavy PBHs. These objects are initially Poisson distributed on scales larger than $r_\cl$, as the perturbation of their number density induced by their discreteness dominates over the small residual correlation,
thus evolving subsequently along what described in Refs.~\cite{Inman:2019wvr,DeLuca:2020jug}, see Fig.~\ref{picture} for a pictorial representation. 

More in detail, according to the  hoop conjecture~\cite{Misner:1973prb}, this happens if
\be
r_\text{\tiny b} \lesssim 2 G M_{\cl},
\ee
which translates into a requirement on the correlation function to be
\be
\label{xiHPBH}
{\rm Hoop}: \quad \Phi \gtrsim
6 \cdot 10^4\, \lp \frac{C}{200}\rp^{-1/6} \lp \frac{r_\cl}{\rm kpc} \rp^{-1}.
\ee
We also expect that, due to frequent BH encounters during the cluster's collapse, strong GW emission may occur, inducing an even more efficient clusteringenesis.

Since PBH clusters follow a Poisson distribution on large scales and they are characterised by a small physical size, they may dynamically evaporate~\cite{1987gady.book.....B} before effectively collapsing into heavy PBHs\footnote{We warn the reader that by evaporation we mean the dissipation of a cluster due to a series of encounters between its constituents. This is not related to the Hawking evaporation that light PBHs experience across their evolution.}. This occurs if the evaporation timescale is smaller than the characteristic free-fall time shown in Eq.~\eqref{freefall}.
The evaporation time of a system of $N_\text{\tiny cl}=M_\text{\tiny cl}/M_\PBH$ PBHs clustered in a region of size $r_\text{\tiny b}$ and subject to the gravitational force is given by~\cite{1987gady.book.....B}
\begin{align}
t_{\text{\tiny{ev}}}
& \simeq 14 \frac{N_\text{\tiny cl}}{\log N_\text{\tiny cl}} \frac{r_\text{\tiny b}}{v_\text{\tiny b}} \nonumber \\
& \simeq \frac{10^{11} {\rm yr}}{\log N_\text{\tiny cl}} \left( \frac{N_\text{\tiny cl}}{10^6} \right)^{1/2} \left(  \frac{M_\PBH}{M_\odot}\right)^{-1/2}  \left( \frac{r_{\text{\tiny b}}}{{\rm pc}} \right)^{3/2}.
\end{align}
By imposing that evaporation is slower than free-fall, $t_{\text{\tiny{ev}}} \gtrsim \tau_\cl$, one then gets
\be
N_\cl \gtrsim 6 \cdot 10^{-6} \xi_0^2 \lp \frac{\tau_\cl}{\rm yr} \rp \lp \frac{C}{200} \rp^{1/2},
\ee
such that using $N_\cl \sim \overline{n}_\PBH r_\cl^3 \xi_0$ one gets the condition to avoid cluster evaporation
\be
\label{xievap}
{\rm Eva}: \quad \Phi \gtrsim 5.5 \cdot 10^{-4}  \lp \frac{M_\PBH}{M_\odot} \rp \lp \frac{r_\cl}{\rm kpc} \rp^{-3}.
\ee
This bound has to be interpreted to be conservative since relaxation may occur on a timescale larger than free-fall, and thus make evaporation even less efficient. Moreover, the initial collapse might produce a BH before the cluster size relaxes to $r_\text{\tiny b}$ with a smaller compactness parameter $C$.

\begin{figure}[t]
	\centering
\includegraphics[width=0.49\textwidth]{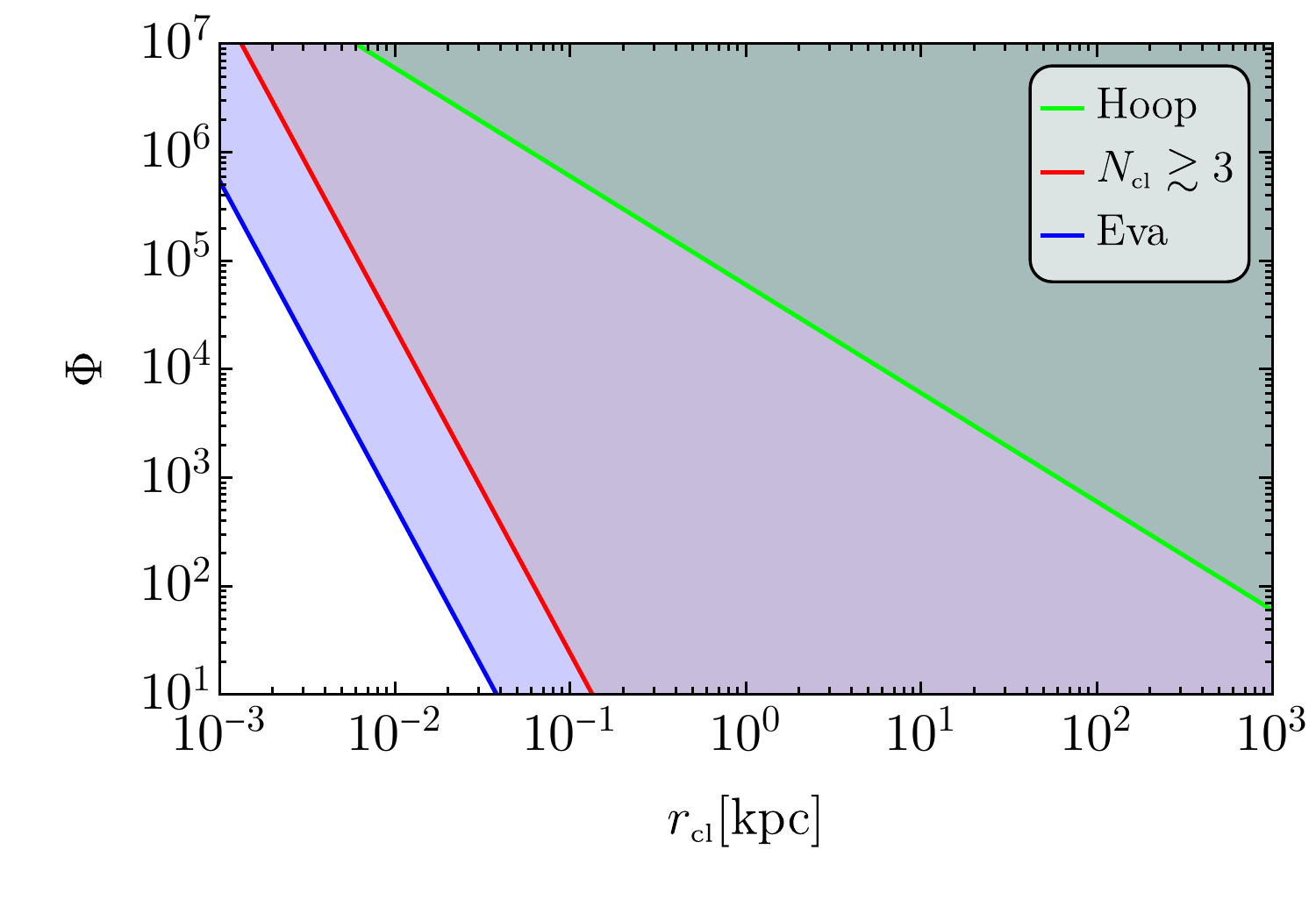}
	\caption{
 Constraints on the allowed parameter space for initial PBH clustering to form heavy PBHs, assuming solar-mass PBHs $M_\PBH = M_\odot$. 
	In red we shade out the region where the condition requiring sufficiently large clusters ($N_\cl \gtrsim 3$, Eq.~\eqref{Ncl3}) is satisfied.
The blue region corresponds to evading cluster evaporation (Eva, Eq.~\eqref{xievap}) while the green one indicates where heavy PBH formation occurs (Hoop, Eq.~\eqref{xiHPBH}).
At clustering scales larger than about Mpc, bounds coming from CMB anisotropies on isocurvature perturbations need to be taken into account, giving the upper bound $\Phi \lesssim 10^{-6}$~\cite{Planck:2018vyg, Young:2015kda, DeLuca:2021hcf}.
	}
	\label{allowedparameterspace}
\end{figure}

By combining this constraint with the requirement of heavy PBHs  formation in Eq.~\eqref{xiHPBH} and assuming that, for definiteness, each PBH cluster contains at least three PBHs~\cite{DeLuca:2022uvz}
\be
\label{Ncl3}
N_\cl \gtrsim 3: \quad \Phi \gtrsim 2.3 \cdot 10^{-2}  \lp \frac{M_\PBH}{M_\odot} \rp
\lp \frac{r_\text{\tiny cl}}{\rm kpc} \rp^{-3}, \,
\ee
one can show in Fig.~\ref{allowedparameterspace} the allowed parameter space on the clustering model to have the formation of heavy PBHs. From the figure it is clear that the bound coming from the hoop conjecture sets the strongest constraint on the heavy PBH mass, which is then found to be at least
\be
M_\HPBH \gtrsim 2.4 \cdot 10^{4} \lp \frac{\Phi}{10^6}\rp^{-2} \lp \frac{C}{200} \rp^{-1/2}  M_\odot .
\ee
It is important to stress that the parameter space of these newly formed heavy PBHs is constrained by the same set of observational bounds that apply to the standard PBHs that are formed in the radiation-dominated epoch, see Ref.~\cite{Carr:2020gox} for a review. 
However, as their formation is induced by different initial conditions with respect to the standard scenario, their relation to indirect probes may change.

\vskip 0.5cm
\noindent
\noindent{{\bf{\it Implications.}}}
The production of heavy PBHs from the gravitational collapse of initially clustered PBH seeds provides a novel scenario for PBH formation and has several implications, some of which we described below.

In the standard scenario where the PBHs are created by the collapse of large overdensities created during inflation when they reenter the Hubble radius in the radiation-dominated phase, there is  a standard correlation between the PBH mass, as a fraction of the mass enclosed in the cosmological horizon at the time of  horizon crossing, and the corresponding comoving momentum $k$~\cite{Sasaki:2018dmp}
\be
\label{standardREl}
M_\PBH \simeq 2.45 \cdot 10^{6} \lp \frac{k}{\rm kpc}\rp^{-2} M_\odot.
\ee
In the clusteringenesis this relation does not hold anymore. 

There are two straightforward implications of this fact. In the standard scenario, the power spectrum of the curvature perturbation to generate PBHs must be of the order of $\mathcal{O}(10^{-2})$. Therefore, using Eq.~\eqref{standardREl}, the current constraints on the $\mu$-distortions  would rule out PBHs in the mass range $(10^4 \divisionsymbol 10^{14})M_\odot$~\cite{Nakama:2017xvq,Byrnes:2018txb,DeLuca:2021hcf}. On the other hand, within our mechanism, PBHs with such masses may be generated through large curvature perturbations at much smaller scales, thus evading the bounds.

The second point is related to the generation of the stochastic gravitational wave background (SGWB) associated to PBH production due to the nonlinear nature of gravity (see Ref.~\cite{Domenech:2021ztg} for a review). In the standard formation scenario of heavy PBHs, one expects a relation between the PBH mass and the peak frequency of the SGWB $f_\SGWB$ as~\cite{Garcia-Bellido:2017aan}
\be
f_\SGWB \simeq 0.4 \lp \frac{M_\HPBH}{M_\odot} \rp^{-1/2} {\rm nHz}.
\ee
On the other hand, when heavy PBHs are produced through the scenario proposed in this work, 
one expects an induced SGWB related to the small scale perturbations inducing the formation of the light PBHs. 
Furthermore, a large spatial correlation at formation time and the dynamics of cluster collapse may eventually induce a GW emission due to the presence of putative time dependent quadrupolar anisotropies \cite{DeLuca:2019llr,Dalianis:2020gup,Domenech:2021and,Flores:2022uzt}, whose characteristic frequency would depend on the clustering parameter space.
Finally, the formation of binaries composed by light PBHs during the cluster collapse is expected to induce the emission of GWs with a spectrum of frequencies determined by the Innermost stable circular orbit $f_\text{\tiny ISCO} \simeq 2 \cdot 10^3{\rm Hz}/M_\text{\tiny PBH}$.
The differences in the spectrum of GWs produced in this scenario might constitute a unique signature of the clusteringenesis.

Another possible application of PBH clusteringenesis  is related to the generation of Supermassive BHs (SMBHs).  Such BHs are believed to sit at the center of galaxies and they have been recently observed at high redshifts $z\gtrsim 6$, providing a challenge to standard formation mechanisms~\cite{Volonteri:2010wz,Volonteri:2021sfo}. 
Furthermore, SMBHs currently account for about $10^{-5}$ the dark matter density in the universe~\cite{Yu:2002sq,Shankar:2007zg}.
Using the  Schechter function to describe the SMBH mass function at high-redshifts, it has been assessed that the mean SMBH mass is approximately of the order of $10^{10} M_\odot$~\cite{Willott:2010yu,Volonteri:2011rm}, which might be the final result after a phase of efficient baryonic mass accretion starting from SMBH seeds with masses $M_\SMBH \approx 10^{4} M_\odot$~\cite{Nakama:2017xvq,Serpico:2020ehh}.

Scenarios involving the presence of PBH seeds able to generate SMBHs invoke an efficient phase of baryonic mass accretion during their cosmological evolution, and require values of the PBH abundance smaller than $f_\PBH \lesssim 10^{-9}$ to avoid bounds coming from CMB data~\cite{Serpico:2020ehh}. Within the clusteringenesis scenario, on the other hand, one can efficiently produce heavy PBHs with masses comparable to a SMBH seed at formation time. In particular, to efficiently give rise to a SMBH seed with mass larger than $M_\SMBH \approx 10^{4} M_\odot$, by inspecting Eq.~\eqref{standardREl} one finds $\Phi \gtrsim 10^6$ or $\xi_0 \gtrsim 10^{15}$ for $f_\PBH \lesssim 10^{-9}$.
Let us also notice that the SMBHs generated with this mechanism, with low abundance and masses larger than about $10^{10} M_\odot$, may accelerate the formation of bright and massive galaxies at very high redshifts, as recently suggested in Ref.~\cite{Liu:2022bvr}. 

\vskip 0.5cm
\noindent
\noindent{{\bf{\it Conclusions.}}}
Given our ignorance of the initial clustering of PBHs, we have shown that heavy PBHs may be generated, even during the radiation phase, during the collapse of a halo composed by previously generated lighter PBHs.
This happens if the hoop conjecture inequality is satisfied, provided that the PBH clusters do not dynamically evaporate away.
We have discussed possible implications of this scenario, in particular providing a new explanation to the generation of the  supermassive black holes observed at the center of galaxies and pointing out that constraints on the curvature perturbation at a given scale may be uncorrelated with the mass of the PBHs. 

It will be interesting to investigate other possible implications of the clusteringenesis scenario, for instance characterising in more detail the mass distribution of the heavy PBHs coming from the collapse, and extending the standard Press-Schechter formalism to initial non-linear densities and thresholds.
Moreover, in our work we have assumed spherical symmetry to perform analytical estimates. It would be important to understand the role of asphericities in the dynamics of the collapse. We expect that angular momentum may potentially reduce the efficiency of the process and  decrease the abundance of heavy PBHs, which could still be consistent with the small abundance necessary for heavy PBH to represent the SMBH seeds. Finally, we stress that a fully relativistic  simulation would be needed to investigate the role of non-linearities and relativistic corrections which may arise at the formation of the heavy BHs. 

\vskip 0.5cm
\noindent
\noindent{{\bf{\it Acknowledgments.}}}
We thank  H. Veermäe for useful and interesting feedback on the draft.
V.DL. is supported by funds provided by the Center for Particle Cosmology at the University of Pennsylvania. 
G.F. acknowledges financial support provided under the European Union's H2020 ERC, Starting Grant agreement no.~DarkGRA--757480 and under the MIUR PRIN programme, and support from the Amaldi Research Center funded by the MIUR program ``Dipartimento di Eccellenza" (CUP:~B81I18001170001). This work was supported by the EU Horizon 2020 Research and Innovation Programme under the Marie Sklodowska-Curie Grant Agreement No. 101007855.
A.R. is supported by the Boninchi Foundation for the project ``PBHs in the Era of GW Astronomy".

\bibliography{draft}

\end{document}